\documentclass[aps,prl,showpacs,showkeys, groupedaddress,twocolumn,superscriptaddress]{revtex4-1}

\usepackage{amsmath,graphicx,latexsym,times,color}
\usepackage{setspace}
\usepackage{hyperref}
\usepackage{array}
\usepackage{titlesec}
\usepackage{physics}
\usepackage{color}
\usepackage[T1]{fontenc}
\usepackage[version=3]{mhchem} 
\newcommand{\parallelsum}{\mathbin{\!/\mkern-5mu/\!}}
\usepackage{empheq} 
\newcommand{\V}[1]{\ensuremath{\mathbf{#1}}} 

\let\oldtimes\times  
\renewcommand\times{{\oldtimes}}

\begin{document}

\author{Dongzhe Li}
\email{dongzhe.li@cemes.fr}
\affiliation{CEMES, Universit\'e de Toulouse, CNRS, 29 rue Jeanne Marvig, F-31055 Toulouse, France}

\author{Soumyajyoti Haldar}
\affiliation{Institute of Theoretical Physics and Astrophysics, University of Kiel, Leibnizstrasse 15, 24098 Kiel, Germany}

\author{Stefan Heinze}
\affiliation{Institute of Theoretical Physics and Astrophysics, University of Kiel, Leibnizstrasse 15, 24098 Kiel, Germany}

\title{Strain-Driven Zero-Field Near-10 nm Skyrmions in Two-Dimensional van der Waals Heterostructures}

\begin{abstract} 
Magnetic skyrmions -- localized chiral spin structures -- show great promise for spintronic applications. The recent discovery of two-dimensional (2D) magnetic materials opened new opportunities for exploring such topological spin structures in atomically thin van der Waals (vdW) materials. Despite recent progress in stabilizing metastable skyrmions in 2D magnets, their diameters are still beyond 100~nm and their lifetime, which is essential for applications, has not been explored yet. Here, using first-principles calculations and atomistic spin simulations, we predict that compressive mechanical strain leads to stabilizing zero-field skyrmions with diameters close to 10 nm in a Fe$_3$GeTe$_2$/germanene vdW heterostructure. The origin of these unique skyrmions is attributed to the high tunability of Dzyaloshinskii-Moriya interaction and magnetocrystalline anisotropy energy by strain, an effect which is shown to be general for Fe$_3$GeTe$_2$ heterostructures with buckled substrates. Based on our first-principles parameters for the magnetic interactions, we calculate the energy barriers protecting skyrmions against annihilation and their lifetimes using transition-state theory. We show that nanoscale skyrmions in strained Fe$_3$GeTe$_2$/germanene can be stable for hours at temperatures up to 20 K.
\end{abstract}


\maketitle
	
Magnetic skyrmions \cite{Bogdanov1989,Bogdanov1994b} have attracted tremendous attention due to their intriguing topological properties and promising applications for next-generation spintronics devices \cite{fert2017magnetic,everschor2018perspective,Luo2018,gobel2021beyond,li2021magnetic,Psaroudaki2021}. The Dzyaloshinskii-Moriya interaction (DMI), which prefers a canting of the spins of adjacent magnetic atoms, is often recognized as the key ingredient in forming such localized noncollinear magnetic structures. The DMI originates from spin-orbit coupling (SOC) and relies on broken inversion symmetry. During the last decade, in order to obtain large interfacial DMI, first observed in ultrathin transition-metal films \cite{bode2007chiral,Ferriani2008,Heinze2011},
	a comprehensive effort has been denoted to develop ferromagnetic/heavy metal (FM/HM) multilayers with strong SOC \cite{moreau2016additive,soumyanarayanan2017,boulle2016room,legrand2020room}.
	
	In 2017, the discovery of truly two-dimensional (2D) magnetic materials \cite{gong2017discovery,huang2017layer,deng2018gate} opened up new opportunities for exploring novel magnetic phenomena in reduced dimensions. In particular, 
	two independent experimental groups reported the first observation of skyrmions in the van der Waals (vdW) magnets Cr$_2$Ge$_2$Te$_6$ \cite{han2019topological} and Fe$_3$GeTe$_2$ \cite{ding2020observation}. After that, several experimental groups observed stable skyrmion lattices in 2D magnets vdW heterostructures such as FGT/WTe$_2$ \cite{wu2020neel}, FGT/Co/Pd multilayer \cite{yang2020creation}, and FGT/Cr$_2$Ge$_2$Te$_6$ \cite{wu2021van}. These exciting measurements have broadened magnetic skyrmion 
interface platforms from conventional transition-metal films and multilayers to vdW materials consisting of weakly bonded 2D layers. Stabilizing skyrmions in 2D magnets has several potential advantages. These include avoiding pinning by defects due to high-quality vdW interfaces, the possibility of forming skyrmions with a minimum thickness (i.e., a single layer limit), and easy control of magnetism via external stimuli such as strain, electric field, light, or magnetic field. Up to now, the reported skyrmion size in 2D vdW heterostructures is 
	above 100 nm \cite{han2019topological,ding2020observation,wu2020neel} at least one order of magnitude larger than the desired diameter ($<$10 nm) required for memory applications \cite{fert2017magnetic}. 

	From the theoretical point of view, during the last three years, atomistic spin models parametrized from first-principles calculations based on 
	density functional theory (DFT) have been extensively used to explore the DMI in 2D magnets. 
	Several strategies have been proposed to achieve DMI in these materials by breaking inversion symmetry. Particularly, the family of Janus vdW magnets, which lacks inversion asymmetry with different atoms occupying top and bottom layers, has been predicted to possess large enough DMI to generate skyrmions \cite{Liang2020,Yuan2020,cui2020strain,Changsong2020,Zhang2020,shen2021strain,Cui_NL2021,du2022spontaneous}. Additionally, it has been shown that when an electric field is applied perpendicular to the CrI$_3$ monolayer, the DMI emerges due to the breaking of inversion symmetry \cite{Liu2018}. Another 
	focus was designing full vdW heterostructures consisting of magnetic and non-magnetic 
	vdW layers, tuning the DMI through the proximity effect \cite{sun2020controlling,sun2021manipulation}. Moreover, metal FM/2D materials interfaces are predicted to possess significant DMI with strong magnetocrystalline anisotropy energy (MAE) \cite{Hallal2021,Shao2022}. 
	The Néel-type magnetic skyrmions observed in Fe$_3$GeTe$_2$ crystals were 
	attributed to the DMI due to the oxidized interfaces \cite{Park2021}. Ferroelectrically controllable skyrmions \cite{CuiPRR2021,Dou2022} 
	have also been recently proposed. However, previous studies were only focused on the
	formation of isolated skyrmions and a quantification of individual skyrmion stability and lifetime in 2D magnets, crucial for all device applications, is so far missing in the literature.
	
	In this Letter, using first-principles calculations and atomistic spin simulations, we demonstrate that a mechanical strain can significantly enhance skyrmion stability and lifetime in 2D vdW heterostructures, leading to 
	{\em zero-field near-10 nm skyrmions}. We focus on Fe$_3$GeTe$_2$ (FGT) vdW heterostructures of current experimental interest with a high Curie temperature of about 230 K, which can be increased up to room temperature by patterning \cite{li2018patterning} and electron doping \cite{deng2018gate}. The origin of these nanoscale skyrmions at zero magnetic fields is attributed to the highly tunable DMI and MAE by strain. We show that the DMI is significantly enhanced by more than 400\% when applying a small compressive strain while the MAE is dramatically reduced to 25\% of its original value. The strained FGT heterostructure also displays a considerable exchange frustration which can increase skyrmion stability~\cite{Malottki2017}.
	We show that the efficient strain-driven DMI and MAE control is general for FGT heterostructures with buckled substrates.
	Finally, we calculate energy barriers between the metastable skyrmion and the ferromagnetic ground state and quantify the stability of
	skyrmions by calculating their lifetimes as a function of magnetic field and temperature. Our work demonstrates the possibility to stabilize skyrmions with diameters on the order of 10~nm down to zero magnetic field in 2D vdW heterostructures.
	
	We performed density functional theory (DFT) calculations using two 
	DFT codes which differ in their choice of basis set: The \textsc{Fleur} code \cite{fleurv26} is based on the full-potential linearized augmented plane wave (FLAPW) formalism while the \textsc{QuantumATK} package \cite{smidstrup2019} uses an expansion of electronic states in a linear combination of atomic orbitals (LCAO). The former ranks among the most accurate implementations of DFT, while the latter is computationally much 
	less demanding, therefore, useful for large-scale systematic or explorative studies.
	For the DMI calculation, two different approaches are used: The generalized Bloch theorem (gBT) \cite{Kurz2004,Heide2009,Zimmermann2019} and the supercell approach \cite{YangPRL_2015}, which are denoted in the following as FLAPW-gBT and LCAO-supercell. Computational details are given in Supporting Information (SI) Section I.

	The total energies for different collinear and non-collinear spin structures obtained via
	DFT are used to parameterize the extended Heisenberg model which is given by:
	\begin{equation}\label{model}
		\begin{split}
			H & =-\sum_{ij}J_{ij}(\V{m}_i \cdot \V{m}_j)-\sum_{ij}\V{D}_{ij} \cdot(\V{m}_i \times \V{m}_j) \\
			& -\sum_i K_i (m_i^z)^2 - \sum_i M(\V{m}_i \cdot B_{\text{ext}})
		\end{split}
	\end{equation}
	where $\V{m}_i$ and $\V{m}_j$ are normalized magnetic moments at position $\V{R}_i$ and $\V{R}_i$ respectively. $M$ denotes the size of the magnetic moment at every site. The four magnetic interaction terms correspond to the Heisenberg isotropic exchange, the DMI, the magnetic anisotropy energy (MAE), and the external magnetic field, and they are characterized by the interaction constants $J_{ij}$, $\V{D}_{ij}$, and $K_i$, and	$B_{\text{ext}}$, respectively. Note that our spin model is adapted to a collective 2D model by treating three Fe layers of the FGT as a whole system, similar to a monolayer system. All magnetic interaction parameters are measured in meV/unit cell (uc).
\begin{figure}[tbp]
	\centering
	\includegraphics[width=1\linewidth]{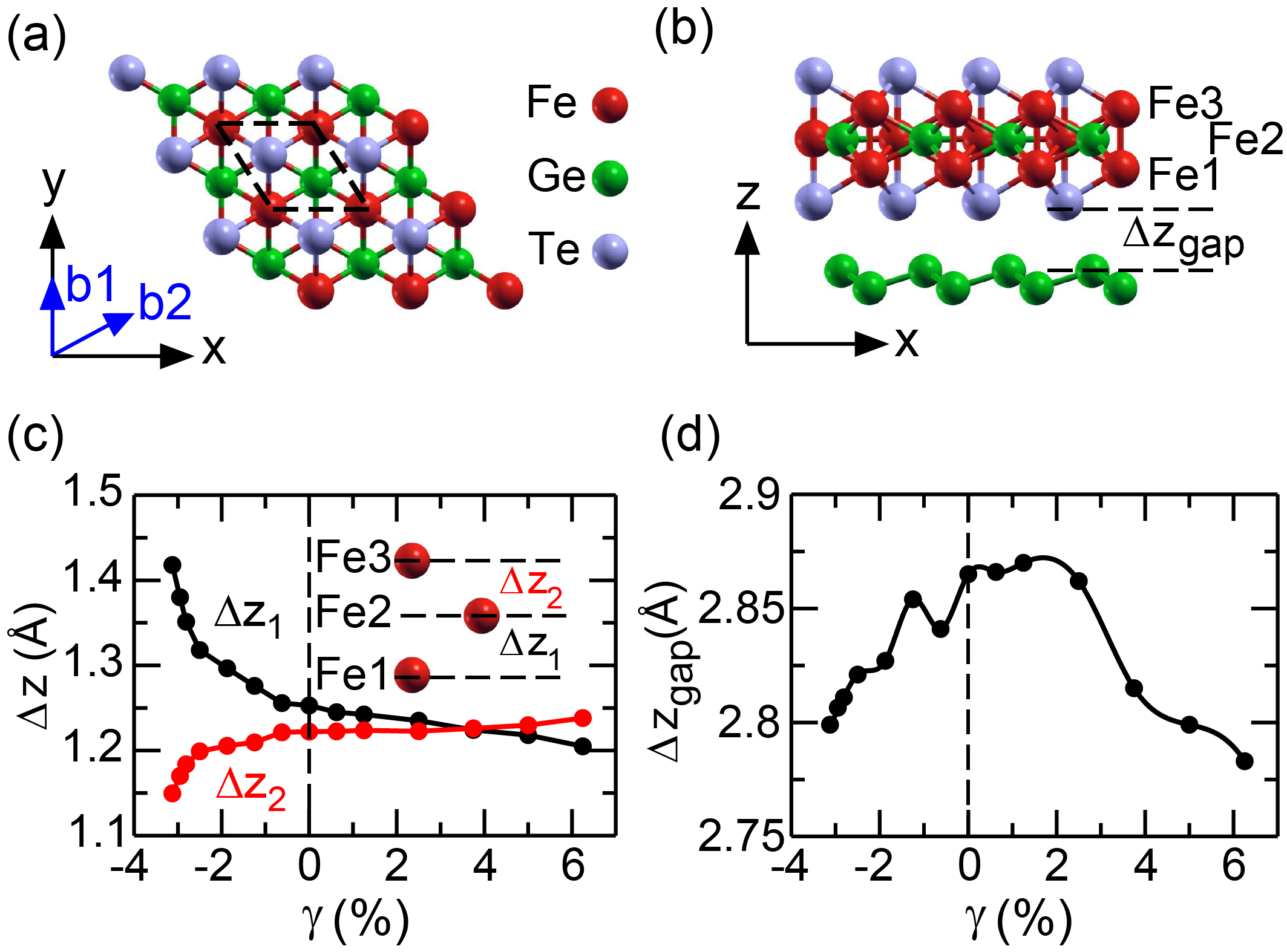}
	\caption{\label{FGT_ge_structure} (a) Top view of the atomic lattice of the Fe$_3$GeTe$_2$ monolayer. The dashed lines denote the 2D unit cell. 
	(b) Side view of the Fe$_3$GeTe$_2$ monolayer on germanene.
	$(x, y, z)$ and (b1, b2) are the Cartesian coordinates and crystallographic directions, respectively. Dependence of (c) structural deformation $\Delta z_1$, $\Delta z_2$ and (d) vdW gap, $\Delta z_{\rm gap}$, on biaxial strain for the FGT/Ge heterostructure.}
\end{figure}
	
	We consider a vdW heterostructure where an FGT monolayer is deposited on germanene (denotes as FGT/Ge in the following). As shown in Fig.~\ref{FGT_ge_structure}(a), the FGT adopts the space groups (194) P6$_3$/\textit{mmc} and can be seen as a stack of three Fe hexagonal lattices with an hcp stacking. In the following, the top, center, and bottom (interface) atoms are denoted as Fe3, Fe2, and Fe1, respectively. For the atomic relaxation, we employed the generalized gradient approximation (GGA), obtaining a relaxed lattice constant of 4.00~\AA~for the FGT monolayer, which is in good agreement with experimental data of about 3.991 $\sim$ 4.03 \AA \cite{Deiseroth2006,chen2013magnetic}. Then, germanene is matched at the interface with the FGT with a lattice mismatch smaller than 1\%. In order to better describe the vdW gap ($\Delta z_{\text{gap}}$), we took into account vdW interactions using semi-empirical dispersion corrections as formulated by Grimme \cite{grimme2010consistent}. For magnetic exchange calculations, we used the local density approximation (LDA) without Hubbard $U$ correction since it yields a magnetic moment of 1.76~$\mu_{\text{B}}$/Fe that compares well with experiments, as also pointed out in Ref. \cite{Zhuang2016,deng2018gate}. The lowest-energy stacking configuration is the one where the Te atom is right above the center of the hexagonal ring of germanene with an optimized vdW gap of about 2.86~\AA~(see Fig.~\ref{FGT_ge_structure}b), which agree well with previous results \cite{Junjie2021}.
	
	\begin{table*}[t]
\caption{Shell-resolved Heisenberg exchange constants ($J_{n}$), Dzyaloshinskii-Moriya interaction constants ($D_{n}$), and magnetocrystalline anisotropy energy ($K$) obtained via DFT for a collective 2D spin model (i.e., three Fe atoms are treated as a whole) of FGT/Ge at $\gamma = 0\%$ and $\gamma = -3\%$. A positive (negative) sign of $D_n$ denotes a preference of CW (CCW) rotating cycloidal spin spirals. A positive (negative) sign of $K$ indicates an easy in-plane (out-of-plane) magnetization direction.} \label{table_dmi}
		\centering
			\begin{tabular}{cccccccccccccccccc}
				\hline\hline
				\multicolumn{1}{c}{$$} & \multicolumn{1}{c} {~~$J_{1}$~~} & \multicolumn{1}{c} {~~$J_{2}$~~} & \multicolumn{1}{c} {~~$J_{3}$~~} & \multicolumn{1}{c} {~~$J_{4}$~~} & \multicolumn{1}{c} {~~$J_{5}$~~} & \multicolumn{1}{c} {~~$J_{6}$~~} & \multicolumn{1}{c} {~~$J_{7}$~~} & \multicolumn{1}{c} {~~$J_{8}$~~} & \multicolumn{1}{c} {~~$D_{1}$~~} & \multicolumn{1}{c} {~~$D_{2}$~~} & \multicolumn{1}{c} {~~$D_{3}$~~} & \multicolumn{1}{c} {~~$D_{4}$~~} & \multicolumn{1}{c} {~~$D_{5}$~~} & \multicolumn{1}{c} {~~$D_{6}$~~} & \multicolumn{1}{c} {~~$D_{7}$~~} & \multicolumn{1}{c} {~~$K$~~}\\ 
				$\gamma=0\%$~ & 22.87 & $-0.21$ & $-1.78$ & 0.43 & 1.25& $-0.31$ & 0.00 & $-0.15$ & 0.23 & 0.09 & 0.00 & 0.00 & 0.00 & 0.00 & 0.00 & $-3.90$ \\
				$\gamma=-3\%$~  & 21.51  & $-2.89$ & 0.82 & 0.40 & $-1.41$ & $-0.88$ & 0.89 & $-0.44$ & 0.97  & 0.00 & $-0.02$ & $-0.06$ & 0.04 & 0.02 & 0.02 & $-0.86$\\
				\hline
		\end{tabular}
	\end{table*}		
	
			\begin{figure}[hbtp]
	\centering
	\includegraphics[width=0.8\linewidth]{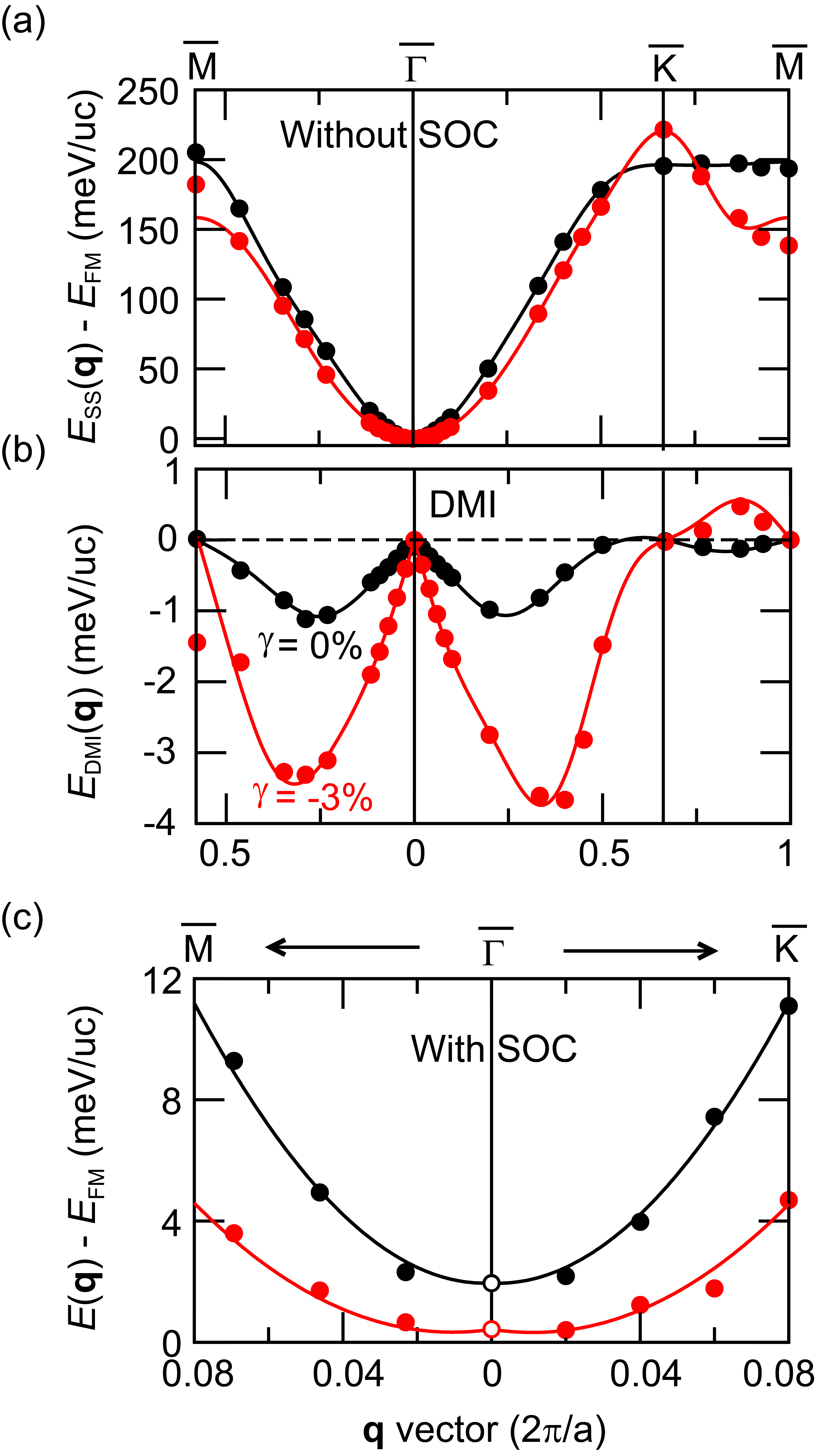}
	\caption{\label{spin-spiral} Strain-dependent energy dispersion of spin spirals for FGT/Ge obtained by FLAPW-gBT. (a) Energy dispersion $E_{\rm SS}(\V{q})$ with respect to the FM state, $E_{\rm FM}$, of planar homogeneous spin spirals for FGT/Ge along the high symmetry path $\overline{\text{M}}$-$\overline{\Gamma}$-$\overline{\text{K}}$-$\overline{\text{M}}$ without SOC. The symbols (black circles for $\gamma=0\%$, red circles for $\gamma=-3\%$) represent the DFT calculations in scalar-relativistic approximation, while the solid lines are the ﬁts to the Heisenberg exchange interaction. (b) The same as (a) but for the energy contribution $E_{\text{DMI}}(\V{q})$ of cycloidal spin spirals due to SOC. Note that positive and negative energies represent a preference of counterclockwise (CCW) and clockwise (CW) spin configurations. (c) Zoom around the FM state ($\overline{\Gamma}$ point), including the Heisenberg exchange, the DMI, and the MAE, i.e.~$E(\V{q})=E_{\rm SS}(\V{q})+E_{\text{DMI}}(\V{q})+K/2$. The DMI leads to the local energy minima for CW rotating spin spirals, and the MAE is responsible for the constant energy shift ($K/2$) of the spin spirals with respect to the FM state.}
\end{figure}
	
	For 2D materials, the lattice, electronic and magnetic properties can be easily modified through interfacing with various substrates or fabricating stretchable heterostructures \cite{gong2019two,jiang2021recent}. This indicates the possibility of tuning magnetic interactions
	and even inducing distinct spin textures in FGT heterostructures through strain engineering. We model the strain by applying the in-plane biaxial strain along the lattice vectors, and we fully relax structures for each strained lattice. The biaxial strain is defined as $\gamma=(a-a_0)/a_0$, where $a$ and $a_0$ are the strained and unstrained lattice constants of FGT heterostructures, respectively. Both compressive and tensile strains are considered ranging from $-3$\% to 6.25\%. 
	It has been demonstrated by \textit{ab initio} phonon spectrum calculations that FGT is stable under such strains~\cite{hu2020enhanced}.
	We emphasize that the value of $a_0=4.0$~\AA~used in this work was evaluated 
	in GGA, slightly higher than the lattice constant calculated 
	in LDA, $a_0=3.91$~\AA. If we use the latter as a reference, the largest compressive strain used in this work becomes less than $-1$\%. 
	
	A remarkable structural modification is observed in Fig.~\ref{FGT_ge_structure}c which shows the strain dependence of vertical, i.e., $z$ component, distances between 
	Fe atoms of the different layers: Fe1-Fe2, $\Delta z_1$, and Fe2-Fe3, $\Delta z_2$
	(see sketches in Figs.~\ref{FGT_ge_structure}b,c). As a negative strain is applied, $\Delta z_1$ increases rapidly, while $\Delta z_2$, in contrast, decreases. Such a 
	large geometrical change is due to the buckled structure of germanene, which introduces sizable atomic forces under negative strains. We find the acting force on the Ge atom (middle layer of FGT) is negative (i.e., towards to germanene layer) while the one on Fe2 is positive. This leads to an increase in structural asymmetry, which might be important for DMI modifications. In addition, the vdW gap in FGT/Ge decreases for both compressive and tensile strains (Fig.~\ref{FGT_ge_structure}d), leading to an increase in hybridization strength at the interface.
	
	Now we turn to the influence of strain on the ground state and magnetic interactions in FGT/Ge.  Fig.~\ref{spin-spiral} shows the calculated energy dispersion $E(\V{q})$ of flat homogeneous spin spirals per unit cell using FLAPW-gBT. We first focus on the results in the scalar-relativistic approximation, i.e.~without SOC 
	(see Fig.~\ref{spin-spiral}a). The energy dispersions are calculated along the high symmetry directions $\overline{\Gamma \text{M}}$ and $\overline{\Gamma \text{KM}}$ of the 2D hexagonal Brillouin zone (BZ). The $\overline{\Gamma}$ point corresponds to the FM state, the $\overline{ \text{M}}$ point to the row-wise AFM state, and the $\overline{ \text{K}}$ point to the Néel state with angles of 120$^{\circ}$ between adjacent spins. 
	
	Both the FGT/Ge interfaces at zero strain ($\gamma = 0\%$) and at a compressive strain 
	($\gamma = -3\%$) possess a minimum of $E(\V{q})$ at the $\overline{\Gamma}$ point, i.e., the FM state, and they exhibit a large FM nearest-neighbor exchange constant, as clearly seen from the large energy difference between the $\overline{\Gamma}$ and $\overline{\text{M}}$ point (AFM state). 
	We note that the strained FGT/Ge heterostructure ($\gamma=-3\%$) displays 
	a stronger exchange frustration which can be seen in Table \ref{table_dmi} from 
	the increased antiferromagnetic exchange constants of beyond nearest neighbor (NN) shells. Therefore, the NN approximation fails at large $\V{q}$ for strained FGT/Ge, as demonstrated in Fig.~S1 in SI.
	
		\begin{figure*}[tp]
		\centering
		\includegraphics[width=0.8\linewidth]{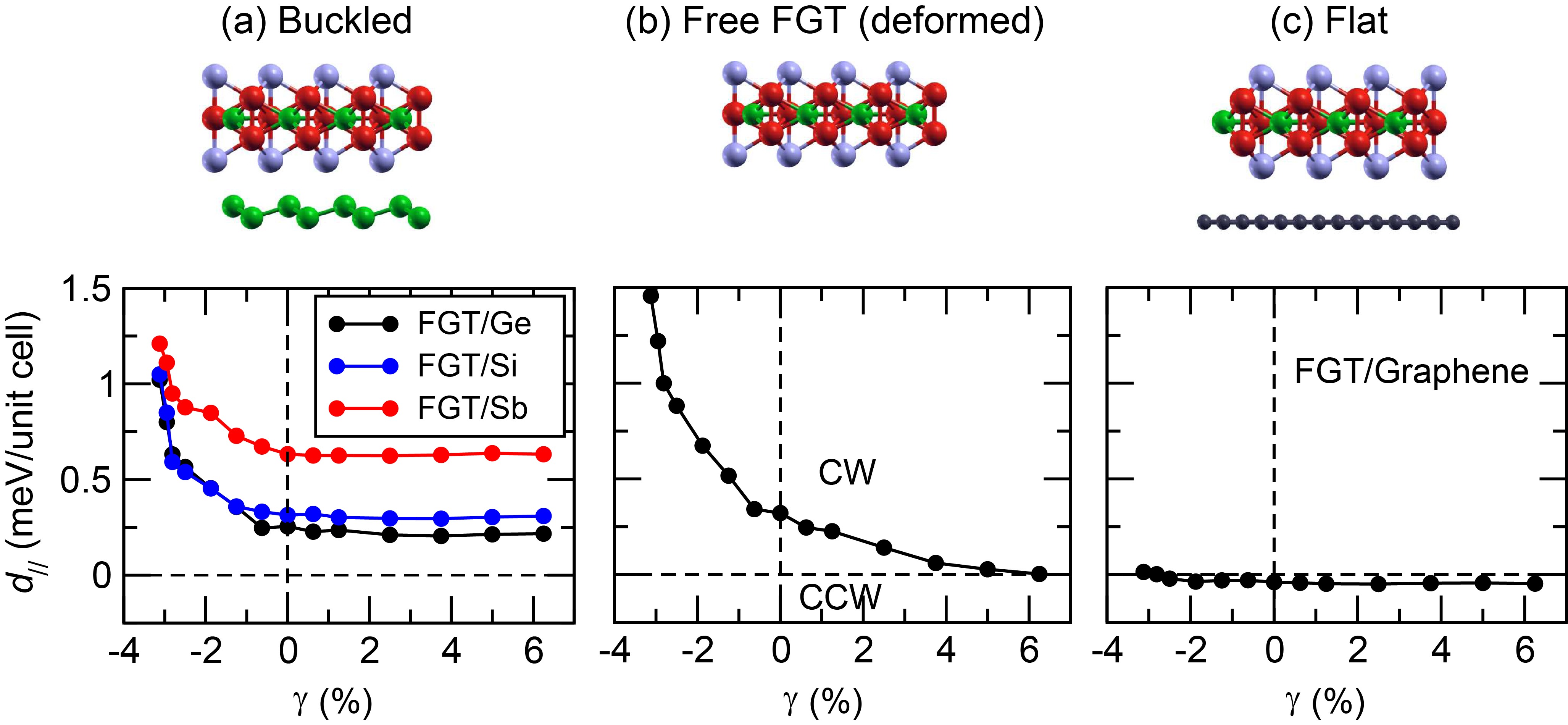}
		\caption{\label{general} (a) Sketch of the structure of a monolayer FGT deposited 
			on buckled 2D vdW substrates and
			calculated in-plane component of DMI as functions of biaxial strain $\gamma$.
			(b) The same as in (a) but for the free-standing FGT with 
			structural deformations as in FGT/Ge. 
			(c) The same as in (a) but for FGT/graphene heterostructure.}
	\end{figure*}

	Upon including SOC, the DMI arises in systems with broken inversion symmetry.
	%
	For free-standing FGT, the DMI involving either Fe1 or Fe3 have opposite signs (i.e., chirality) because of the (001) mirror plane \cite{Laref2020}. Upon incorporating germanene in FGT, the inversion symmetry breaking at the FGT/Ge interface gives rise to an emergent DMI. The DMI favors cycloidal spiraling magnetic structures with a unique rotational sense.
	For FGT/Ge a clockwise rotational sense is favored as seen from the calculated negative energy contribution due to SOC, $E_{\rm DMI}(\V{q})$, to the dispersion of cycloidal spin spirals 
	(Fig.~\ref{spin-spiral}b).
	In the vicinity of the $\overline{\Gamma}$ point an isotropic DMI, expected for 
	a system with $C_{3v}$ symmetry, is apparent from the 
	same slope in $\overline{\Gamma \text{M}}$ and 
	$\overline{\Gamma \text{K}}$ directions.
	
The DMI constants obtained from a fit of the spin model, Eq.~(1), to the DFT energies (Table \ref{table_dmi}) are dominated by the NN term, $D_1$, and the values rapidly decrease with distance. This indicates that the NN approximation fits very well in FGT heterostructures. According to Moriya’s rule \cite{Moriya1960}, the $\V{D}_{ij}$ can be expressed as $\V{D}_{ij}=d_{\parallelsum}~(\hat{\V{u}}_{ij} \times \hat{\V{z}})+d_{\perp}\hat{\V{z}}$, where $\hat{\V{u}}_{ij}$ being the unit vector between sites $i$ and $j$ and $\hat{\V{z}}$ indicating normal to the plane. It has been shown that the out-of-plane component of $\V{D}_{ij}$ plays a negligible role in forming skyrmions in 2D magnets \cite{Liang2020,du2022spontaneous}. Therefore, we focus on its in-plane component DMI, $d_{\parallelsum}$. The $d_{\parallelsum}$ (also known as microscopic DMI coefficient) can be approximated by $D_1$ fitted to an effective NN DMI model presented in Fig.~S1 in SI. The micromagnetic DMI coefficient $D$ can be calculated using
   \begin{equation}\label{supercell_M}
	D=\frac{3\sqrt2d_{\parallelsum}}{N_{\text{F}}a^2}
\end{equation}
where $a$ and $N_{\text{F}}$ are the lattice constant and the number of ferromagnetic layers, respectively.

In the equilibrium case ($\gamma = 0\%$), we find a preference for clockwise rotating spin spirals with $d_{\parallelsum}=0.25$~meV. The corresponding micromagnetic DMI coefficient, is about $|D|$ = 0.36 mJ/m$^{2}$. This value is comparable to the other two FGT heterostructures reported recently as promising interfaces to host stable skyrmion states:
FGT/In$_2$Se$_3$ \cite{Huang2022} ($\sim$0.28 mJ/m$^{2}$) and FGT/Cr$_2$Ge$_2$Te$_6$ \cite{wu2021van} ($\sim$0.31 mJ/m$^{2}$).
In the strained heterostructure ($\gamma=-3\%$) the DMI increases by a factor of about 4
to a value of $d_{\parallelsum}=1$~meV.
Such a large DMI is comparable to state-of-the-art FM/HM interfaces such as Co/Pt (1.47 meV) \cite{Zimmermann2019}, Pd/Fe/Ir(111) (1.0 meV) \cite{dupe2014tailoring},
and Fe/Ir(111) (1.7 meV) \cite{Heinze2011}. Its corresponding $D$ is 1.52 mJ/m$^{2}$, which is more than one order of magnitude higher than the critical DMI value of 0.1 mJ/m$^{2}$ \cite{wu2020neel} necessary for stabilizing skyrmions in FGT heterostructures. 
This indicates a possibility of strain control of magnetic skyrmions in these systems. In contrast, a tensile strain 
($\gamma>0$) has almost no effect on the DMI.

	The inclusion of SOC also induces MAE. 
	The easy magnetization axis of FGT/Ge is out-of-plane and the values of the MAE are $K = -3.90$ and $-0.86$ meV/unit cell at $\gamma=0\%$ and $\gamma=-3\%$, respectively, i.e.~the compressive strain decreases the MAE to about 25\% of its equilibrium value.
	The energy contribution from MAE leads to an energy offset of $K/2$ for spin spirals with respect to the FM state as can be seen in a zoom of $E(\V{q})$ 
	around $\overline{\Gamma}$ for cycloidal spin spirals (Fig.~\ref{spin-spiral}c).
	
	The FM state at zero energy is the magnetic
	ground state for unstrained as well as for strained FGT/Ge (Fig.~\ref{spin-spiral}c).
	However, the effect of all interactions, i.e., the Heisenberg exchange, the DMI, and MAE, upon applying a compressive strain is significant. The large reduction of
	the MAE
	shifts the energy dispersion close to the FM state.
	In the regime of small $\arrowvert \V{q} \arrowvert$, 
	the dispersion for $\gamma = 0\%$ shows an energy rise with $\V{q}^2$ while $E(\V{q})$ becomes extremely flat at $\gamma = -3\%$. This indicates considerable
	exchange frustration in strained FGT/Ge which opens the
	possibility of metastable nanoscale skyrmions at zero magnetic field as reported in transition-metal ultra-thin films \cite{meyer2019isolated}.
	
	\begin{figure}[tp]
		\centering
		\includegraphics[width=1\linewidth]{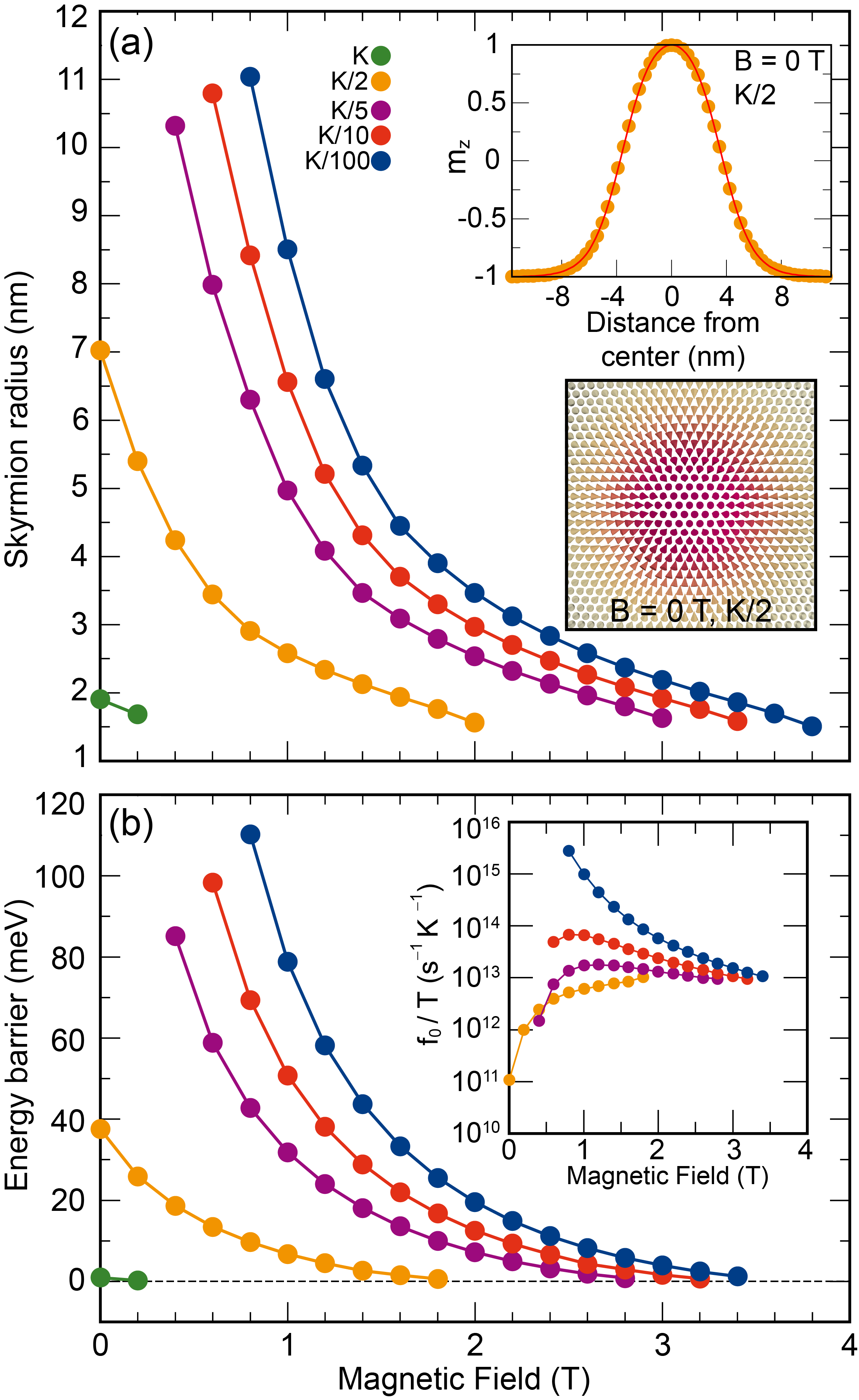}
		\caption{\label{GNEB} Skyrmion radius and energy barrier for skyrmion collapse in 
	strained FGT/Ge ($\gamma = -3\%$) evaluated using magnetic interaction parameters from DFT. (a) Skyrmion radius as a function of the applied magnetic field. The skyrmion profile and spin texture at $B = 0$~T and $K/2$ are shown as insets. (b) Energy barriers of isolated skyrmions versus magnetic field strength. Calculated attempt frequencies $f_0$ are shown on a logarithmic scale in the inset. Both radii and energy barriers are evaluated at different MAE values, namely $K$ (green), $K$/2 (yellow), $K$/5 (purple), $K$/10 (red), and $K$/100 (blue), where $K=-0.86$~meV is the MAE estimated by DFT.}
	\end{figure}
	
	To demonstrate that the proposed strain tuning of DMI and MAE is quite general, we performed systematic studies for two other FGT heterostructures with buckled substrate: 
	Fe$_3$GeTe$_2$/silicene (FGT/Si) and Fe$_3$GeTe$_2$/antimonene (FGT/Sb). These calculations have been done using the LCAO-supercell approach since it is 
	computationally 
	less demanding than FLAPW-gBT. 
	The LCAO-supercell is in excellent agreement with FLAPW-gBT for the DMI in FGT/Ge 
	(see Fig.~S2 in SI). Fig.~\ref{general}a shows the variation of DMI under applied 
	strain for three FGT interfaces. 
	Clearly the significant enhancement of DMI for a small compressive strain is a robust effect which appears for all three interfaces with buckled vdW substrates. Moreover, the amplitude of DMI for FGT/Sb is much stronger than for the other two interfaces and
	exhibits a value of about 0.65~meV, even in the unstrained condition. 
	We attribute this large increase to the much larger SOC constant of Sb
	which scales approximately as 
	$Z^2$, where $Z$ denotes the nuclear charge. 
	
	When the FGT monolayer is deposited on a substrate, two sources modify its magnetic properties: (i) the structural deformation of the FGT and (ii) the coupling of FGT states to those of the substrate. To remove artificially the latter, we 
	have calculated the DMI for free-standing FGT but distorted as after deposition on germanene (Fig.~\ref{general}b). We recover a similar curve as in the case of FGT/Ge (Fig.~\ref{general}a), confirming that the enhancement of DMI is mainly attributed to the structural deformation and not to interface hybridization. In contrast, when a flat vdW material such as graphene is used as the substrate, the DMI in FGT is almost constant with respect to strain (Fig.~\ref{general}c). Note that due to the weak SOC effect in graphene, we find a tiny DMI of about $-0.04$~meV.
	
	\begin{figure*}[tp]
		\centering
		\includegraphics[width=1\textwidth]{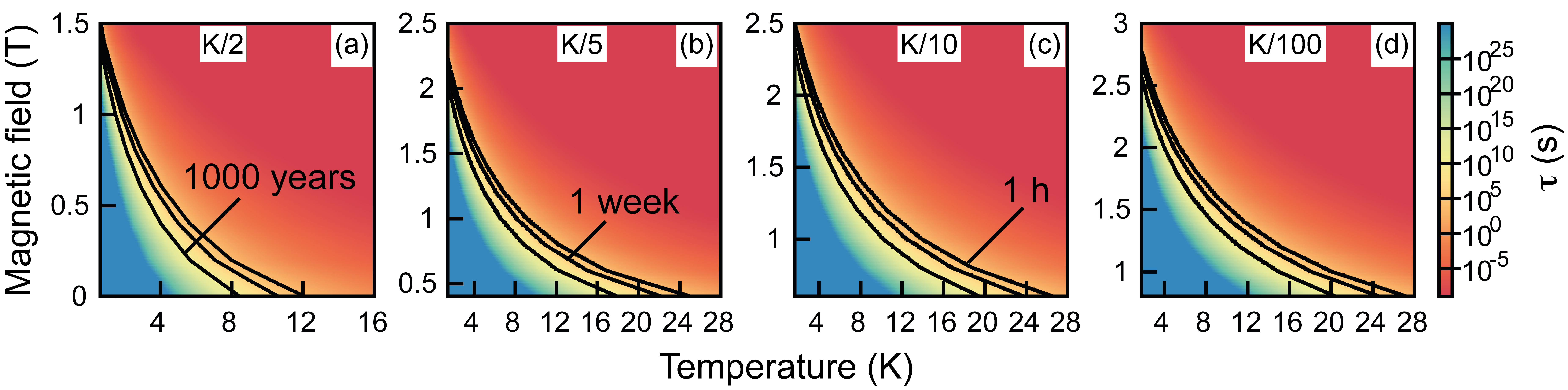}
		\caption{\label{lifetime} The lifetime of skyrmions, $\tau$, in strained FGT/Ge  obtained in harmonic transition state theory based on the spin model
		with DFT parameters at a MAE of (a) $K/2$, (b) $K/5$, (c) $K/10$, and (d) $K/100$ as a function of magnetic field and temperature.}
	\end{figure*}
	
	Since the flat spin spiral energy dispersion with SOC in FGT/Ge under strain (Fig.~\ref{spin-spiral}c) indicates the possibility of stabilizing small magnetic skyrmions down to zero magnetic field we performed atomistic spin simulations using the spin model described by Eq.~\ref{model} with the full set of DFT parameters. Isolated skyrmions were obtained using spin dynamics. Minimum energy paths for the skyrmion collapse processes into the ferromagnetic ground state were identified using the geodesic nudged elastic band (GNEB) method~\cite{bessarab2015method} and the collapse barriers for skyrmions were obtained from the saddle point along the path. Finally, we used harmonic transition state theory (HTST) to quantify skyrmion stability by calculating their lifetime~\cite{bessarab2018lifetime,Haldar2018,Malottki2019,muckel2021experimental} (for computational details see SI).
	
	In our simulations, we create isolated skyrmions in the FM background based on the theoretical profile given in Ref.~\cite{Bogdanov1994b} and fully relax these spin structures by solving the damped Landau-Lifschitz equation self-consistently. We find no magnetic skyrmion emerging in FGT/Ge without applying strain ($\gamma=0\%$). This is due to a very large out-of-plane MAE ($K$) with a moderate DMI amplitude. Interestingly, when a small strain is applied ($\gamma=-3\%$), our spin dynamics simulations predict N\'eel-type skyrmions with a diameter (= 2 $\times$ radius) of about 4~nm stabilized in the FM background at zero magnetic field (Fig.~\ref{GNEB}a). Since the MAE of FGT can be greatly suppressed by doping or temperature \cite{tan2018hard,wang2020modifications,Park2020_nl}, 
	we present in 
	Fig.~\ref{GNEB}a the skyrmion radius with respect to the applied magnetic field under different values of the MAE. For skyrmion profiles at different MAE values, please refer to Fig. S3.
	
	Remarkably, nanoscale skyrmions with radii ranging from 2 nm to 11 nm are observed at small magnetic fields of less than 0.5 T and the radii rapidly decrease with applied magnetic field as expected. In particular, at $K$ and $K/2$, small skyrmions with radii of about 2 nm and 7 nm are obtained at zero magnetic field ($B = 0$~T, 
	see inset of Fig.~\ref{GNEB}a for skyrmion profile at $K/2$ and Fig.~S3 for profile at $K$ and $K/10$). When a small magnetic field is applied ($B < 0.5$~T), we obtain a small size skyrmion within sub-10 nm. Such small skyrmions are technologically desired for future skyrmionic devices. In strained FGT/Ge, they originate from the extremely flat spin spiral curve close to the FM state (Fig.~\ref{spin-spiral}c). The interplay of exchange and DMI reduces the energy cost of a fast spin rotation while large MAE enforces the fast spin rotation. Note that all previously reported magnetic skyrmions in 2D magnets were stabilized under finite magnetic fields. 
	
	To get insight into the stability of nanoscale skyrmions in FGT/Ge, we have calculated the minimum energy path for a transition between the metastable skyrmion and the FM ground state using the GNEB method~\cite{bessarab2015method}. Surprisingly, for a $\text{MAE}$ of $K$, although the 2 nm radius skyrmion can be stabilized at zero-field in spin dynamics simulations, its energy barrier is almost zero (Fig.~\ref{GNEB}b), i.e.~it is actually unstable. This highlights the importance of going beyond conventional spin dynamics
	to establish stability of topological spin structures.
	
	For a $\text{MAE}$ of $K/2$, we already obtain a significant 
	energy barrier of around 40~meV protecting skyrmions from collapse  
	at $B = 0$~T (Fig.~\ref{GNEB}b). Upon further decreasing the MAE the energy barrier
	is further enhanced. The 
	skyrmion is annihilated via the radial symmetric mechanism
	in which the skyrmion shrinks symmetrically to the saddle point 
	(SP) and then collapses into the FM state~\cite{muckel2021experimental}. 
	As expected the DMI protects the skyrmion state while the MAE prefers the FM state and decreases the total barrier (Fig.~S4 in SI). This immediately explains that the energy barrier increases upon decreasing the MAE. A second effect is an increase of the DMI contribution due to the increased skyrmion radius at reduced MAE. The exchange frustration is not as large as in ultrathin transition-metal film systems~\cite{Malottki2017,meyer2019isolated}, but it provides only a small negative contribution to the barrier (see Fig.~S4 in SI).
	
The stability of skyrmions can be quantified by their lifetime, $\tau$,
which is given by the Arrhenius law $\tau = f_{0}^{-1} \exp(\frac{\Delta E}{k_{\V{B}} T})$, where $\Delta E$, $f_{0}$, and $T$ are energy barrier,
attempt frequency, and temperature, respectively.
$f_{0}$, obtained within harmonic transition state 
theory~\cite{bessarab2018lifetime},
depends strongly on magnetic field and on the value of the MAE 
(inset of Fig.~\ref{GNEB}b). This effect is similar to that
observed in ultrathin transition-metal films which can be traced back to a
change of entropy with skyrmion radius 
and profile~\cite{Malottki2019,varentcova2020toward}.

From the temperature and field dependence of the skyrmion lifetime (Fig.~\ref{lifetime}) we predict that isolated skyrmions in strained FGT/Ge are stable up to hours at a temperature ranging from 12 $\sim$ 28 K 
down to zero field depending on the MAE. Therefore, these skyrmions can be probed, e.g., by scanning tunneling microscopy or Lorentz transmission electron microscopy experiments. 

To summarize, we predict the formation of nanoscale -- close to 10 nm diameter -- skyrmions in 2D vdW heterostructures 
down to
zero magnetic field induced by mechanical strain. In particular, our DFT calculations demonstrate that a small compressive strain can enhance the DMI up to 400\% while decreasing the MAE dramatically down to 25\% of its value in Fe$_3$GeTe$_2$/Ge. Notably, such highly tunable magnetic interactions turn out to be general for FGT heterostructures with buckled substrates (e.g., silicene, antimonene). Using atomistic spin simulations, we further explicitly calculate skyrmion energy barriers as well as lifetimes. We demonstrate that near-10 nm are stable down to zero magnetic field with lifetimes of hours up to 20~K. Our findings thus show a route to realize sub-10 nm skyrmions in 2D vdW heterostructures at zero magnetic field.


\section*{Acknowledgement}
	We thank Gustav Bihlmayer for valuable discussions. This study has been (partially) supported through the grant NanoX n° ANR-17-EURE-0009 in the framework of the "Programme des Investissements d'Avenir". We gratefully acknowledge financial support from the Deutsche Forschungsgemeinschaft (DFG, German Research Foundation) through SPP2137 "Skyrmionics" (project no.~462602351). Numerical calculations were performed using HPC resources from CALMIP (Grant 2022-[P21023]).

\section*{}
The Supporting Information is available free of charge at \url{https://pubs.acs.org/}. Computational details on DFT and atomistic spin simulations, effective nearest-neighbor model, comparison between the LCAO-supercell and FLAPW-gBT approaches, and minimum energy paths of skyrmion collapse with different energy decomposition.

	\bibliography{test}

\end{document}